# Exact analytical PGSE signal for diffusion confined to a cylindrical surface using a spectral Laplacian formalism

Erick J Canales-Rodríguez[1,2,3,*], Chantal M.W. Tax[4,5], Juan Manuel Górriz[1,2,3], Derek K. Jones[5],

Jean-Philippe Thiran[6,7,8], Jonathan Rafael-Patiño[6,7]

[1] Department of Signal Theory, Networking and Communications, University of Granada, Granada, Spain

[2] Andalusian Research Institute in Data Science and Computational Intelligence (DaSCI), University of Granada, Spain

[3] Research Centre for Information and Communication Technologies (CITIC), University of Granada, Granada, Spain

[4] Image Sciences Institute, University Medical Center Utrecht, The Netherlands.

[5] Cardiff University Brain Research Imaging Centre (CUBRIC), Cardiff University, Cardiff, Wales, United Kingdom.

[6] Signal Processing Laboratory 5 (LTS5), Ecole Polytechnique Fédérale de Lausanne (EPFL), Lausanne, Switzerland.

[7] Computational Medical Imaging & Machine Learning Section, Center for Biomedical Imaging (CIBM), Lausanne, Switzerland

[8] Department of Radiology, Centre Hospitalier Universitaire Vaudois (CHUV), Lausanne, Switzerland

* Corresponding author: ejcanalesr@ugr.es

**Abstract**

*Objective.* To derive an exact analytical expression for the diffusion MRI signal arising from diffusion confined to a cylindrical surface under finite rectangular pulsed-gradient spin-echo (PGSE) gradients, and to develop computationally efficient strategies for repeated model evaluations. *Approach.* The Bloch–Torrey equation was solved using a spectral matrix formalism of the Laplace operator in the eigenbasis of the cylindrical surface. The resulting signal is expressed as a product of non-commuting matrix exponentials and is valid for arbitrary rectangular gradient durations and separations, without approximations to the diffusion propagator or spin phase distribution. A reduced real spectral basis was introduced to decrease the dimensionality of the problem. Accelerated implementations based on Strang splitting and Gauss–Legendre quadrature were developed for repeated signal evaluations and computation of the spherical mean. The analytical signal was validated against Monte Carlo diffusion simulations. *Main results.* The exact formulation showed excellent agreement with Monte Carlo simulations over a wide range of cylinder radii and acquisition parameters. The spectral expansion converged rapidly with mode order, and the reduced basis substantially decreased computational cost without loss of accuracy. The accelerated implementations provided large additional speed-ups while retaining controllable errors relative to the exact solution. *Significance.* The proposed framework extends exact finite-pulse solutions of the Bloch–Torrey equation to diffusion confined to cylindrical surfaces and provides practical implementations for directional and orientationally averaged diffusion MRI signals. It therefore offers an approximation-free reference for assessing faster signal approximations and for applications requiring large numbers of model evaluations.

## 1. Introduction

Magnetic Resonance Imaging (MRI) experiments sensitized to molecular self-diffusion provide a powerful framework to probe transport processes in confined environments and to infer geometric features of restricting boundaries in porous media and biological tissues (Mitra and Sen 1992, Le Bihan 2007, Brownstein and Tarr 1979, Basser *et al* 1994). In pulsed-gradient spin-echo (PGSE) experiments (Stejskal and Tanner 1965), diffusion sensitization arises from the accumulation of spin phase under time-dependent magnetic field gradients, such that the measured macroscopic signal reflects ensemble-averaged properties of the underlying stochastic motion. Establishing rigorous links between the measured signal and the geometry of confinement remains a central problem in the theory of diffusion MRI (dMRI) (Novikov *et al* 2018, Alexander *et al* 2019, Veraart *et al* 2020, Jones *et al* 2018, Alexander *et al* 2010, Palombo *et al* 2020, Jelescu *et al* 2020, Özarslan *et al* 2009, Grebenkov 2010, Nilsson *et al* 2017, Dyrby *et al* 2018, Canales-Rodríguez *et al* 2024, Barakovic *et al* 2023).

From a theoretical perspective, the dMRI signal from water molecules in a bounded domain can be described exactly by solving the Bloch–Torrey equation. Spectral and matrix-based approaches employing the eigenfunctions of the Laplace operator have been developed to address this problem (Robertson 1966, Caprihan *et al* 1996, Callaghan 1997, Callaghan and Codd 1998, Codd and Callaghan 1999, Barzykin 1999, Axelrod and Sen 2001, Grebenkov 2006, 2008a, 2007). These frameworks provide a rigorous description of restricted diffusion and, in principle, allow one to treat arbitrary gradient waveforms and boundary conditions (Grebenkov 2008b, 2009). However, for specific geometries of interest, practical signal models often adopt simplifying assumptions (most notably the narrow-pulse limit, approximate treatments of finite gradient durations (Lori *et al* 2003), or the Gaussian-phase approximation (van Gelderen *et al* 1994, Balinov *et al* 1993, Söderman and Jönsson 1995)) in order to obtain expressions that are analytically tractable and computationally efficient.

Diffusion confined to cylindrical surfaces constitutes a particularly relevant case. In biological tissues, this geometry has been proposed as an effective model for myelin water diffusion, in which water molecules are constrained to move along the surface of the myelin sheath surrounding axons (Canales-Rodríguez *et al* 2025, Peled *et al* 1999). This assumption is supported by the ultrastructure of myelin, where adjacent lamellae are separated by narrow aqueous gaps with an effective thickness of approximately 3–4 nm, as inferred from diffraction- and scattering-based reconstructions (Inouye and Kirschner 2016, Inouye *et al* 2017). In a previous study, we derived an exact analytical expression for the PGSE signal arising from diffusion confined to a cylindrical surface in the narrow-pulse limit and proposed an approximate extension to account for finite gradient durations (Canales-Rodríguez *et al* 2025). While that formulation showed good agreement with Monte Carlo simulations over a broad range of experimental conditions, noticeable discrepancies emerged for large diffusivities and relatively high diffusion weightings, highlighting the limitations of the approximations when finite gradient pulse effects become significant.

The main objective of the present work is to resolve this limitation by deriving an exact analytical solution of the Bloch–Torrey equation for diffusion taking place on the two-dimensional cylindrical manifold, without invoking any approximation on the diffusion propagator, resulting spin phase distribution, or gradient pulse duration. The derivation is based on the spectral matrix formalism of the Laplace operator (Grebenkov 2008b, 2009), specifically constructed for the cylindrical surface geometry. By explicitly accounting for the non-commutativity of the operators involved during the different segments of the PGSE sequence, we obtain a closed-form dMRI signal expression valid for arbitrary combinations of rectangular gradient duration and separation.

The paper is organized as follows. **Section 2** outlines the theoretical framework for describing confined diffusion in PGSE experiments using the Laplacian spectral matrix formalism. In **Section 3**, this framework is applied to diffusion confined to a cylindrical surface, and the resulting analytical signal expression is derived. Building on this exact formulation, we also develop efficient numerical strategies for repeated signal evaluations, including a Strang splitting approximation and an efficient computation of the spherical mean signal using Gauss–Legendre quadrature. Details of the Monte Carlo diffusion simulations used for numerical validation are also provided. **Section 4** presents the comparison between analytical and Monte Carlo signals and benchmarks the accelerated implementations against the exact formulation to quantify accuracy–runtime trade-offs. The scope and limitations of the proposed approach are discussed in **Section 5**. Additional technical details of the mathematical derivations are provided in the **Appendices**.

## 2 Theory

In this section, we briefly recall the theoretical framework for describing confined diffusion based on the spectral decomposition of the Laplace operator. The use of Laplace operator eigenfunctions to solve the Bloch–Torrey equation was originally introduced by Robertson (Robertson 1966), and subsequently extended through a variety of analytical and numerical approaches. Notable developments include the multiple narrow-pulse approximation (Caprihan *et al* 1996), matrix-based formulations enabling efficient numerical implementations (Callaghan 1997, Callaghan and Codd 1998), and additional spectral treatments (Barzykin 1998, 1999). A more complete theoretical description was later formulated by Axelrod and Sen (Axelrod and Sen 2001) and further generalized and reformulated within the multiple correlation function framework by Grebenkov (Grebenkov 2007). In what follows, we summarize the key relations of this spectral matrix approach that are required for the present work, before applying it to diffusion confined to a cylindrical surface.

### 2.1 Bloch–Torrey equation and spectral approach

The Bloch-Torrey equation in the rotating frame of reference, describing the time evolution of the complex-valued transverse magnetization $M(\mathbf{r},t)$, arising from spins diffusing with diffusion coefficient $D$ in a confined/bounded domain $\Omega$ is (Torrey 1956)

$$\frac{\partial M(\mathbf{r},t)}{\partial t} = D\nabla^2_{\Omega,\mathrm{B}} M(\mathbf{r},t) - i\gamma G(t)(\mathbf{g}\cdot\mathbf{r})M(\mathbf{r},t), \tag{1}$$

where $G(t)$ is the magnitude of the applied time-dependent linear magnetic field gradient pointing along the unit vector $\mathbf{g}$, $\mathbf{r}$ is the spin's spatial position vector, $\gamma$ is the gyromagnetic ratio of the nuclei, and $\nabla^2_{\Omega,\mathrm{B}}$ denotes the Laplace *operator* $\nabla^2$ on the domain $\mathbf{r}\in\Omega$ where the diffusion process takes place, which depends on the boundary conditions describing how the spins interact with the boundaries (e.g., $\mathrm{B}$ ={Neumann, Dirichlet, Robin}).

In practice, it is convenient to represent $M(\mathbf{r},t)$ in the orthonormal basis of eigenfunctions $\{\phi_n\}$ resulting from the spectral decomposition of the Laplace operator $\nabla^2_{\Omega,\mathrm{B}}$. These complex-valued eigenfunctions are determined by the following eigenvalue problem (Grebenkov 2008b):

$$-\nabla^2_{\Omega,\mathrm{B}}\phi_n(\mathbf{r}) = \lambda_n\phi_n(\mathbf{r}), \tag{2}$$

where, depending on the geometry of the problem, $n$ belongs either to the set of non-negative integers $n \in \mathbb{N} = \{0,1,\cdots\infty\}$ or to the set of integers $n \in \mathbb{Z} = \{-\infty,\cdots,-1,0,1,\cdots\infty\}$, and the orthonormal eigenfunctions satisfy the property:

$$\int_\Omega \phi_n(\mathbf{r})\phi_m^*(\mathbf{r})d\mathbf{r} = \delta_{n,m}, \tag{3}$$

where $\phi_m^*$ is the complex-conjugate of $\phi_m$, and $\delta_{nm}$ is the Kronecker delta, equal to 1 if $n = m$ and 0 otherwise. For the standard diffusion boundary conditions considered here, the eigenvalues are non-negative, $\lambda_n \geq 0$, regardless of the indexing convention. This follows directly from Eq. (2) after multiplication by $\phi_n^*$ and integration by parts (Grebenkov 2008b).

In this basis, the magnetization can be expanded as:

$$M(\mathbf{r},t) = \sum_n c_n(t)\phi_n(\mathbf{r}), \tag{4}$$

where $c_n(t)$ are time-dependent coefficients to be determined.

Substituting Eq. (4) into Eq. (1), and using the definition in Eq. (2), we obtain

$$\sum_n \frac{\partial c_n(t)}{\partial t}\phi_n(\mathbf{r}) = -D\sum_n c_n(t)\lambda_n\phi_n(\mathbf{r}) - i\gamma G(t)(\mathbf{g}\cdot\mathbf{r})\sum_n c_n(t)\phi_n(\mathbf{r}). \tag{5}$$

Multiplying both sides of Eq. (5) by $\phi_m^*$, integrating over the spatial domain, and using the orthonormality relation in Eq. (3), we obtain

$$\frac{\partial c_m(t)}{\partial t} = -Dc_m(t)\lambda_m - i\gamma G(t)\sum_n c_n(t)\int_\Omega \phi_n(\mathbf{r})(\mathbf{g}\cdot\mathbf{r})\phi_m^*(\mathbf{r})d\mathbf{r}. \tag{6}$$

Since the mode index $m$ is arbitrary, the above equation holds for every spectral mode. Collecting the time-dependent coefficients $c_m(t)$ for all $m$ into the vector $\mathbf{c}(t)$ gives the matrix system

$$\frac{\partial \mathbf{c}(t)}{\partial t} = -D\mathbf{\Lambda}\mathbf{c}(t) - i\gamma G(t)\mathbf{B}\mathbf{c}(t), \tag{7}$$

where $\mathbf{\Lambda}$ is a diagonal matrix of the eigenvalues obtained from Eq. (2): when $n \in \mathbb{N}$, $\mathbf{c}(t) = [c_0(t),\ c_1(t),\ \cdots]^T$ and $\mathbf{\Lambda} = diag([\lambda_0,\ \lambda_1,\ \cdots])$; when $n \in \mathbb{Z}$, $\mathbf{c}(t) = [\cdots,\ c_{-1}(t),\ c_0(t),\ c_1(t),\ \cdots]^T$ and $\mathbf{\Lambda} = diag([\cdots,\ \lambda_{-1},\ \lambda_0,\ \lambda_1,\ \cdots])$. The elements of the coupling matrix $\mathbf{B}$ are given by

$$B_{mn} = \int_\Omega \phi_n(\mathbf{r})(\mathbf{g}\cdot\mathbf{r})\phi_m^*(\mathbf{r})d\mathbf{r}. \tag{8}$$

The solution to the system of differential equations in Eq. (7) is given by (Barzykin 1998, 1999)

$$\mathbf{c}(t) = \Gamma e^{-\int_0^T (D\mathbf{\Lambda} + i\gamma G(t)\mathbf{B})dt} \mathbf{c}(0), \tag{9}$$

where $\Gamma e$ denotes the time-ordered matrix exponential and $\mathbf{c}(0)$ is the vector evaluated at time zero. In next sections we will determine matrices $\mathbf{\Lambda}$ and $\mathbf{B}$ for the confining domain under study, but first, we will relate the dMRI signal with the time dependent vector $\mathbf{c}(t)$ and the eigenfunctions in the next subsection.

### 2.2 Diffusion MRI signal: Expansion on the spectral basis

The measured dMRI signal is obtained by spatially integrating the transverse magnetization, which can be represented in the basis of eigenfunctions introduced in the previous section:

$$\begin{aligned} E(t) &= \int_\Omega M(\mathbf{r},t)d\mathbf{r} \\ &= \sum_n c_n(t)\int_\Omega \phi_n(\mathbf{r})d\mathbf{r} \\ &= \mathbf{w}\cdot\mathbf{c}(t), \end{aligned} \tag{10}$$

where $\mathbf{w}$ is the vector with entries equal to the spatial integral of the eigenfunctions: $\mathbf{w} = \left[\int_\Omega \phi_0(\mathbf{r})d\mathbf{r},\ \int_\Omega \phi_1(\mathbf{r})d\mathbf{r},\ \ldots\right]$ when $n \in \mathbb{N}$, or $\mathbf{w} = \left[\ldots,\ \int_\Omega \phi_{-1}(\mathbf{r})d\mathbf{r},\ \int_\Omega \phi_0(\mathbf{r})d\mathbf{r},\ \int_\Omega \phi_1(\mathbf{r})d\mathbf{r},\ \ldots\right]$ when $n \in \mathbb{Z}$. This expression establishes a direct relationship between the dMRI signal and vector $\mathbf{c}(t)$.

### 2.3 Analytical solution for the PGSE experiment

In a PGSE spin-echo sequence, the effect of the 180° refocusing pulse can be incorporated into an effective diffusion-encoding gradient waveform, in which gradients applied after the refocusing pulse contribute with the opposite sign to the accumulated phase. Therefore, for two identical rectangular diffusion-gradient pulses of duration $\delta$, separated by $\Delta - \delta$, the effective gradient function can be represented as piecewise constant in three segments (Stejskal and Tanner 1965):

1. First pulse: $G(t) = G$ for $t \in [0, \delta]$,
2. Free evolution: $G(t) = 0$ for $t \in (\delta, \Delta]$,
3. Second pulse: $G(t) = -G$ for $t \in (\Delta, \Delta + \delta]$.

For this sequence, the general solution given in Eq. (9) becomes (Barzykin 1998, 1999, Grebenkov 2008b)

$$\begin{aligned} \mathbf{c}(t) &= e^{-\int_\Delta^{\Delta+\delta} (D\mathbf{\Lambda} + i\gamma G(t)\mathbf{B})dt} e^{-\int_\delta^\Delta (D\mathbf{\Lambda})dt} e^{-\int_0^\delta (D\mathbf{\Lambda} + i\gamma G(t)\mathbf{B})dt} \mathbf{c}(0), \\ &= e^{-(D\mathbf{\Lambda} - i\gamma G\mathbf{B})\delta} e^{-D(\Delta-\delta)\mathbf{\Lambda}} e^{-(D\mathbf{\Lambda} + i\gamma G\mathbf{B})\delta} \mathbf{c}(0). \end{aligned} \tag{11}$$

It should be noted that this relationship involves matrix exponentials applied to square matrices, which are not equal to the ordinary exponential function. As the involved matrices do not commute, we cannot regroup them into a single exponential. The numerical evaluation must be performed by multiplying the exponential matrices from right to left: propagating the effect of the first gradient pulse on the initial state, then including the effect of the free evolution part on the resulting vector, and finally the effect of the second pulse. Nevertheless, as matrix $\mathbf{\Lambda}$ is diagonal, the matrix exponential in the second term in Eq. (11) can be efficiently computed by exponentiating each entry on the main diagonal.

## 3. Methods

### 3.1 Diffusion confined on cylindrical surfaces: Computing matrices $\mathbf{\Lambda}$ and $\mathbf{B}$

After having reviewed the general theory of the spectral approach and resulting dMRI signal, the last missing ingredient of the recipe is to compute the matrices $\mathbf{\Lambda}$ and $\mathbf{B}$ for our problem of interest: for diffusion confined to a cylindrical surface.

In this section, we will focus on deriving an expression for the dMRI signal $E_{\perp}$ for a diffusion gradient applied perpendicularly to the cylinder's main axis, which thus, depends on the spin's displacements on the 2D circumference of the circle perpendicular to this axis.

In polar coordinates, the Laplace operator for this geometry reduces to its angular component part:

$$\nabla^2_{\Omega,\mathrm{B}} = \frac{1}{r^2}\frac{\partial^2}{\partial\theta^2}, \tag{12}$$

where $\theta$ is the azimuthal angle and $r$ is the cylinder/circle radius.

The normalized eigenfunctions of $\nabla^2_{\Omega,\mathrm{B}}$ are (Griffiths and Schroeter 2018)

$$\phi_n = \frac{1}{\sqrt{2\pi}} e^{in\theta},\ n \in \mathbb{Z} = \{-\infty, \cdots, -1, 0, 1, \cdots \infty\} . \tag{13}$$

By plugging Eq. (13) into Eq. (12) we obtain

$$-\nabla^2_{\Omega,\mathrm{B}}\phi_n = \frac{1}{r^2}\frac{\partial^2\phi_n}{\partial\theta^2} = \frac{n^2}{r^2}\frac{1}{\sqrt{2\pi}}e^{in\theta} = \frac{n^2}{r^2}\phi_n . \tag{14}$$

Thus, the eigenvalues are given by (see Eq. (2)) $\lambda_n = \frac{n^2}{r^2}$, which are used to build matrix $\mathbf{\Lambda}$:

$$\begin{aligned}\mathbf{\Lambda} &= \frac{1}{r^2}\mathbf{\Theta},\\ \mathbf{\Theta} &= diag\left(\left[\cdots,\ \ n^2,\ \ \cdots,\ \ 1,\ \ 0,\ \ 1,\ \ \cdots,\ \ n^2,\ \ \cdots\right]\right).\end{aligned} \tag{15}$$

Note that the eigenvalues are symmetric around the ground mode $n = 0$ (i.e., $\lambda_{-n} = \lambda_n$, with $\lambda_0 = 0$), and that the ´spatial/angular´ integrals of these eigenfunctions are zero, except for the ground mode:

$$\begin{aligned}\int_0^{2\pi}\phi_n(\theta)d\theta &= \frac{1}{\sqrt{2\pi}}\int_0^{2\pi}e^{in\theta}d\theta \\ &= \frac{1}{\sqrt{2\pi}}\left(\int_0^{2\pi}\cos(n\theta)d\theta + i\int_0^{2\pi}\sin(n\theta)d\theta\right) \\ &= \begin{cases}\sqrt{2\pi}, & n=0,\\ 0, & n\neq 0.\end{cases}\end{aligned} \tag{16}$$

For this geometry, the elements of matrix $B_{nm}$ (see Eq. (8); all the details are presented in **Appendix A**) are:

$$\begin{aligned}B_{mn} &= \left(\frac{1}{2\pi}\right)r\int_0^{2\pi}\cos(\theta)e^{in\theta}e^{-im\theta}d\theta, \\ &= \frac{r}{2}\left(\delta_{n,m-1}+\delta_{n,m+1}\right).\end{aligned} \tag{17}$$

This matrix can be rewritten as

$$\mathbf{B} = \frac{r}{2}\mathbf{K}\,, \tag{18}$$

where $\mathbf{K}$ is a tri-diagonal matrix with 1 on the off-diagonals, $K_{n,n-1} = K_{n,n+1} = 1$, and 0 on the main diagonal, $K_{n,n} = 0$:

$$\mathbf{K} = \begin{bmatrix} 0 & 1 & & & \\ 1 & 0 & 1 & & \\ & 1 & \ddots & \ddots & \\ & & \ddots & \ddots & 1 \\ & & & 1 & 0 \end{bmatrix}. \tag{19}$$

### 3.2 Optimizing for dimensionality and speed: Recomputing $\mathbf{\Lambda}$ and $\mathbf{B}$

In the previous section we derived the expressions for $\mathbf{\Lambda}$ and $\mathbf{B}$ for the basis of eigenfunctions. In practice, it is necessary to truncate the basis up to a maximum order $n = M$, resulting in matrices with dimensions of $(2M+1)\times(2M+1)$, including the ground eigenmode. Since evaluating the matrix exponential terms in Eq. (11) is computationally demanding when $M$ is large, a more practical alternative is to build a new compact basis (i.e., $n \in \mathbb{N}$ ) with the same number of eigenfunctions $n = M$, but based on matrices with dimensions of $(M+1)\times(M+1)$. This reduction is particularly advantageous from a computational point of view. Modern algorithms for evaluating the matrix exponential, as implemented in *MATLAB* and *SciPy*, rely on a scaling-and-squaring strategy combined with Padé rational approximations and are dominated by dense matrix–matrix products, whose computational cost scales cubically with the matrix dimension (Higham 2006, Al-Mohy and Higham 2009). Therefore, the proposed reduction leads to an expected computational gain proportional to $\left((2M+1)/(M+1)\right)^3$. This corresponds to an ideal speed-up factor of eight in the asymptotic regime of very large $M$.

As demonstrated in **Appendix B**, the spectral coefficients associated with modes of opposite order remain identical throughout the evolution, $c_n(t) = c_{-n}(t)$. Consequently, the contributions from the $n$ and $-n$ modes to the magnetization can be combined into a single basis function. For the present diffusion model, this symmetry is preserved for an arbitrary gradient amplitude $G(t)$ along a fixed transverse direction. It may be broken, for example, by circumferential coherent motion or by a time-varying transverse gradient direction, neither of which is considered here.

To preserve the orthonormality convention adopted in **Section 2**, we define the normalized combination

$$\begin{aligned}\tilde{\phi}_n &= \frac{1}{\sqrt{2}}\left(\phi_n + \phi_{-n}\right) \\ &= \frac{1}{2}\sqrt{\frac{1}{\pi}}e^{in\theta} + \frac{1}{2}\sqrt{\frac{1}{\pi}}e^{-in\theta} \\ &= \sqrt{\frac{1}{\pi}}\cos(n\theta).\end{aligned} \tag{20}$$

For $n > 0$, the two original eigenfunctions are orthogonal and individually normalized; therefore, their direct sum has squared norm 2, which accounts for the factor $1/\sqrt{2}$ in Eq. (20). The ground mode $n = 0$ has no distinct counterpart and is therefore left unchanged.

Hence, the new basis to represent the magnetization, as in Eq. (4), is defined as

$$\tilde{\phi}_n = \begin{cases} \dfrac{1}{\sqrt{2\pi}}, & n = 0, \\ \sqrt{\dfrac{1}{\pi}}\cos(n\theta), & n \geq 1. \end{cases} \tag{21}$$

The resulting compact basis remains orthonormal, satisfying

$$\int_0^{2\pi} \tilde{\phi}_n(\theta)\tilde{\phi}_m(\theta)\,d\theta = \delta_{n,m}\,, \tag{22}$$

as follows directly from the standard orthogonality relations of the cosine functions:

$$\int_0^{2\pi} \cos(n\theta)\cos(m\theta)\,d\theta = \begin{cases} 2\pi, & n = m = 0, \\ \pi, & n = m \neq 0, \\ 0, & n \neq m. \end{cases} \tag{23}$$

Having constructed the compact orthonormal basis, the spectral coefficient vector can be transformed from the original $(2M+1)$-dimensional representation to the $(M+1)$-dimensional representation as:

$$\mathbf{c} = \mathbf{T}\tilde{\mathbf{c}}\,, \tag{24}$$

where the $(2M+1)\times(M+1)$ transformation matrix is:

$$\mathbf{T}=\begin{bmatrix} 0 & 0 & 0 & 1/\sqrt{2} \\ \vdots & \vdots & \iddots & \vdots \\ 0 & 1/\sqrt{2} & \cdots & 0 \\ 1 & 0 & \cdots & 0 \\ 0 & 1/\sqrt{2} & \cdots & 0 \\ \vdots & \vdots & \ddots & \vdots \\ 0 & 0 & 0 & 1/\sqrt{2} \end{bmatrix}. \tag{25}$$

By using this transformation, we rewrite Eq. (11) in terms of the vector $\tilde{\mathbf{c}}$ as

$$\tilde{\mathbf{c}}(t)=\mathbf{T}^{T}e^{-(D\mathbf{\Lambda}-i\gamma G\mathbf{B})\delta}e^{-D(\Delta-\delta)\mathbf{\Lambda}}e^{-(D\mathbf{\Lambda}+i\gamma G\mathbf{B})\delta}\mathbf{T}\tilde{\mathbf{c}}(0). \tag{26}$$

Notably, in **Appendix C**, we prove the following property, which follows from the specific structure of $\mathbf{T}$ and of the matrices $\mathbf{\Lambda}$ and $\mathbf{B}$:

$$\mathbf{T}^{T}e^{-\mathbf{A}_3}e^{-\mathbf{A}_2}e^{-\mathbf{A}_1}\mathbf{T}=e^{-\mathbf{T}^{T}\mathbf{A}_3\mathbf{T}}e^{-\mathbf{T}^{T}\mathbf{A}_2\mathbf{T}}e^{-\mathbf{T}^{T}\mathbf{A}_1\mathbf{T}}. \tag{27}$$

Therefore, Eq. (26) can be rewritten as:

$$\tilde{\mathbf{c}}(t)=e^{-(D\tilde{\mathbf{\Lambda}}-i\gamma G\tilde{\mathbf{B}})\delta}e^{-D(\Delta-\delta)\tilde{\mathbf{\Lambda}}}e^{-(D\tilde{\mathbf{\Lambda}}+i\gamma G\tilde{\mathbf{B}})\delta}\tilde{\mathbf{c}}(0), \tag{28}$$

where

$$\begin{aligned}\tilde{\mathbf{\Lambda}}&=\mathbf{T}^{T}\mathbf{\Lambda}\mathbf{T},\\ &=diag\left(\left[\lambda_0, \quad \lambda_1, \quad \cdots\right]\right),\\ &=\frac{1}{r^2}\tilde{\mathbf{\Theta}}, \text{ where } \tilde{\mathbf{\Theta}}=diag\left(\left[0, \quad 1, \quad \cdots, \quad n^2, \quad \cdots\right]\right),\end{aligned} \tag{29}$$

and

$$\begin{aligned}\tilde{\mathbf{B}}&=\mathbf{T}^{T}\mathbf{B}\mathbf{T},\\ &=\frac{r}{2}\tilde{\mathbf{K}}, \text{ where}\end{aligned} \tag{30}$$

$$\tilde{\mathbf{K}}=\begin{bmatrix} 0 & \sqrt{2} & & & \\ \sqrt{2} & 0 & 1 & & \\ & 1 & \ddots & \ddots & \\ & & \ddots & \ddots & 1 \\ & & & 1 & 0 \end{bmatrix}$$

The main benefit of working with the compact basis, Eqs. (28)-(30), is that now the matrix exponentials are evaluated on matrices with reduced dimensionality, i.e., $(M+1)\times(M+1)$ instead of $(2M+1)\times(2M+1)$ in the original basis, allowing for a more efficient implementation.

### 3.3 Diffusion confined on cylindrical surfaces: Computing $\mathbf{w}$ and $\mathbf{c}(0)$

According to our previous results, the dMRI signal $E_{\perp}$ is given by (see Eqs. (10), (28)-(30))

$$E_{\perp}(G,\Delta,\delta)=\mathbf{w}\cdot\left[e^{-\left(\frac{D}{r^2}\tilde{\mathbf{\Theta}}-i\frac{\gamma Gr}{2}\tilde{\mathbf{K}}\right)\delta}e^{-\frac{D(\Delta-\delta)}{r^2}\tilde{\mathbf{\Theta}}}e^{-\left(\frac{D}{r^2}\tilde{\mathbf{\Theta}}+i\frac{\gamma Gr}{2}\tilde{\mathbf{K}}\right)\delta}\tilde{\mathbf{c}}(0)\right], \tag{31}$$

where $\tilde{\mathbf{\Theta}}$ and $\tilde{\mathbf{K}}$ are defined in Eqs. (29) and (30).

However, to evaluate this expression we still need to determine $\mathbf{w}$ from the eigenfunctions (Eq. (21)), and $\mathbf{c}(0)$ from the initial conditions.

From Eq. (16) we see that the spatial integrals in the diffusion geometry (on the circle´s circumference) are zero $w_n=0$, except for the element corresponding to $n=0$: $\mathbf{w}=[w_0, \ 0, \ \cdots]$. This happens because when the ground eigenfunction $\tilde{\phi}_0(\mathbf{r})$ is constant/spatially uniform, like in our problem, the spatial integrals of all the other eigenfunctions are zero due to the orthogonality property, see Eq. (22).

On the other hand, vector $\mathbf{c}(0)$ can be determined from the initial conditions, that is, from the value of $M(\mathbf{r},t=0)$. Assuming a spatially uniform initial magnetization on the cylinder surface implies $c_n(0)=0$ for all $n\neq 0$, i.e., $\mathbf{c}(0)=[1, \ 0, \ \cdots]^T$. This follows from the expansion in Eq. (4) for $t=0$, together with the observation that the only spatially constant eigenfunction is the ground state $\tilde{\phi}_0(\mathbf{r})$.

Thus, Eq. (31) becomes

$$E_{\perp}(G,\Delta,\delta)=w_0\cdot\left[e^{-\left(\frac{D}{r^2}\tilde{\mathbf{\Theta}}-i\frac{\gamma Gr}{2}\tilde{\mathbf{K}}\right)\delta}e^{-\frac{D(\Delta-\delta)}{r^2}\tilde{\mathbf{\Theta}}}e^{-\left(\frac{D}{r^2}\tilde{\mathbf{\Theta}}+i\frac{\gamma Gr}{2}\tilde{\mathbf{K}}\right)\delta}\right]_{0,0}, \tag{32}$$

where we symbolically evaluated the vector multiplications. The subscript in the previous equation indicates that after computing the product of the three matrix exponentials we only take the entry at index $(0,0)$, i.e., the first matrix element corresponding to ground mode, the index-term $n=0$.

Note that if we repeat the PGSE experiment for $G=0$, the above expression is simplified to

$$\begin{aligned}E_{\perp}(G=0,\Delta,\delta)&=w_0\left[e^{-\frac{D(\Delta+\delta)}{r^2}\tilde{\mathbf{\Theta}}}\right]_{0,0},\\&=w_0 e^{-\frac{D(\Delta+\delta)}{r^2}\tilde{\Theta}_{0,0}},\\&=w_0,\end{aligned} \tag{33}$$

where the matrix-exponential becomes the ordinary exponential of the resulting diagonal matrix, and the first element of $\tilde{\Theta}_{0,0} = 0$. This happens because in problems where $\tilde{\phi}_0(\mathbf{r})$ is a constant, like in our case, $\lambda_0 = 0$ is the only eigenvalue that fulfills the equality given in Eq. (2), $\nabla^2_{\Omega,\mathrm{B}}\tilde{\phi}_0(\mathbf{r}) = 0 = \lambda_0\tilde{\phi}_0$.

Thus, Eq. (32) can be written in the more common normalized form:

$$\frac{E_\perp(q,\Delta,\delta)}{E_\perp(q=0)} = \left[ e^{-\frac{D\delta}{r^2}\tilde{\mathbf{\Theta}}+i\frac{qr}{2}\tilde{\mathbf{K}}} e^{-\frac{D(\Delta-\delta)}{r^2}\tilde{\mathbf{\Theta}}} e^{-\frac{D\delta}{r^2}\tilde{\mathbf{\Theta}}-i\frac{qr}{2}\tilde{\mathbf{K}}} \right]_{0,0} \tag{34}$$

which is dimensionless and the normalized signal lies in the unit interval [0,1]. Note that we introduced the magnitude of the q-vector as $q = |\mathbf{q}| = \gamma G\delta$.

This relationship provides an exact analytical expression for the dMRI signal arising from diffusing spins in a cylindrical surface, when the gradient is applied perpendicularly to the cylinder´s axis. Notably, it is not based on any simplifying assumptions about the diffusion propagator, the duration of the gradient pulses, or the distribution of phases. The only approximation involved is numerical, resulting from truncating the number of eigenfunctions and eigenvalues employed to practically evaluate the expression (i.e., the dimensionality of matrices $\tilde{\mathbf{\Theta}}$ and $\tilde{\mathbf{K}}$). Nevertheless, in practice, it usually converges with a few terms as the eigenvalues grow as $n^2$.

### 3.4 Exact reduction to a single matrix exponential via conjugation symmetry

A direct evaluation of Eq. (34) requires two matrix exponentials per signal sample (e.g., gradient direction or $b$-value). An additional reduction in computational cost follows from noting that the first and third matrix exponentials in Eq. (34) (corresponding to the time intervals in which the diffusion gradients are applied in the PGSE experiment) are complex conjugates of each other. Therefore, in practice, it is sufficient to evaluate only one matrix exponential, while the second one can be obtained by complex conjugation. This strategy reduces the signal evaluation computational cost by approximately a factor of two. Note that this additional computational reduction relies on the two effective PGSE gradient lobes having equal magnitude and opposite sign; for a different waveform, the corresponding matrix exponentials must in general be evaluated independently.

### 3.5 Decoupled diffusive motions and dMRI signal components

In the previous section we derived the analytical dMRI signal expression (Eq. (34)) for the case when the diffusion-encoding gradient is applied perpendicularly to the main cylinder´s axis. For an arbitrary gradient orientation, this expression is still valid but some modifications must be implemented.

For an infinitely long cylinder, the diffusion process of spin particles parallel and perpendicular to the cylinder’s main axis is statistically independent. As a result, the dMRI signal can be expressed as the product of the signals arising from displacements parallel and perpendicular to the cylinder’s axis (Assaf *et al* 2004):

$$E(\mathbf{q},\Delta,\delta) = E_\perp(\mathbf{q}_\perp,\Delta,\delta)E_\parallel(\mathbf{q}_\parallel,\Delta,\delta), \tag{35}$$

where $\mathbf{q}_{\|}$ and $\mathbf{q}_{\perp}$ are the projection of the q-space vector $\mathbf{q} = q\hat{\mathbf{g}}$ along the directions parallel and perpendicular to the cylinder's axis, $\mathbf{q} = \mathbf{q}_{\|} + \mathbf{q}_{\perp}$; thus their magnitudes are given by $q_{\|} = |\mathbf{q}_{\|}| = q\cos(\beta)$ and $q_{\perp} = |\mathbf{q}_{\perp}| = q\sin(\beta)$, where $\beta$ is the angle between the diffusion gradient orientation unit vector $\hat{\mathbf{g}}$ and the unit vector along the main cylinder's $\hat{\mathbf{v}}$. A general detailed derivation of this type of decoupled signal model is provided in (Assaf *et al* 2004).

Since the motion of particles along the cylinder's main axis is unrestricted, we assume 1D Gaussian diffusion with a characteristic myelin water diffusivity $D$ on the cylinder's surface. The normalized dMRI PGSE signal $E_{\|}(\mathbf{q}_{\|}, \Delta, \delta)$ arising from these Gaussian displacements is

$$\frac{E_{\|}(\mathbf{q}_{\|}, \Delta, \delta)}{E_{\|}(\mathbf{q}_{\|} = 0)} = e^{-q^2(\Delta - \delta/3)D\cos(\beta)^2}, \tag{36}$$

where $b = q^2(\Delta - \delta/3)$ is the $b$*-value*.

For a cylindrical surface of finite axial length $L$, with reflecting boundary conditions at its two axial ends, the angular and axial motions remain separable. In this case, the unrestricted Gaussian factor in Eq. (36) can be replaced by the corresponding signal for one-dimensional restricted diffusion on a finite interval with reflecting boundaries.

On the other hand, in the expression for $E_{\perp}(\mathbf{q}_{\perp}, \Delta, \delta)$ given by Eq. (34) we must replace $q$ by its projection $q_{\perp}$ along the plane perpendicular to the cylinder axis. Therefore, the final signal expression in Eq. (35) for an arbitrary gradient orientation becomes

$$\frac{E(q, \hat{\mathbf{g}}, \Delta, \delta)}{E(q=0)} = e^{-q^2(\Delta-\delta/3)D(\hat{\mathbf{g}}\cdot\hat{\mathbf{v}})^2} \left[ e^{-\frac{D\delta}{r^2}\tilde{\mathbf{\Theta}} + i\frac{qr\sqrt{1-(\hat{\mathbf{g}}\cdot\hat{\mathbf{v}})^2}}{2}\tilde{\mathbf{K}}} e^{-\frac{D(\Delta-\delta)}{r^2}\tilde{\mathbf{\Theta}}} e^{-\frac{D\delta}{r^2}\tilde{\mathbf{\Theta}} - i\frac{qr\sqrt{1-(\hat{\mathbf{g}}\cdot\hat{\mathbf{v}})^2}}{2}\tilde{\mathbf{K}}} \right]_{0,0}, \tag{37}$$

where we substituted the following trigonometric relationships, $\cos(\beta)^2 = (\hat{\mathbf{g}}\cdot\hat{\mathbf{v}})^2$ and $\sin(\beta) = \sqrt{1-(\hat{\mathbf{g}}\cdot\hat{\mathbf{v}})^2}$, and made the signal to explicitly depend on the experimental (arguments) parameters of the PGSE sequence.

### 3.6 Strang splitting with *p* substeps

While in the previous sections we introduced theoretical developments and implementation aspects to accelerate the evaluation of the radial signal component, additional speed-ups are desirable. This is relevant, for example, when a large number of signals must be evaluated repeatedly across diffusion gradient orientations, or across different radii when the signal model is used within a nonlinear fitting procedure (e.g., to estimate the radius from measured data as an inverse problem).

In this section, we employ the second-order Strang splitting (or Strang–Marchuk operator splitting) method (Strang 1968, Marchuk 1968) to obtain an approximate signal, which can significantly accelerate the signal evaluation while providing explicit control of the approximation accuracy. This approach is widely used in different fields, including the simulation of Hamiltonian systems, quantum mechanics (for approximating the time evolution governed by the time-dependent Schrödinger equation), as well as in fluid dynamics and numerical solutions of partial differential equations.

The idea behind this method is to approximate the matrix exponentials in Eq. (37) as

$$e^{-\frac{D\delta}{r^2}\tilde{\mathbf{\Theta}} \pm i\frac{qr\sqrt{1-(\hat{\mathbf{g}}\cdot\hat{\mathbf{v}})^2}}{2}\tilde{\mathbf{K}}} \approx \left( e^{-\frac{D\delta}{2pr^2}\tilde{\mathbf{\Theta}}} e^{\pm i\frac{qr\sqrt{1-(\hat{\mathbf{g}}\cdot\hat{\mathbf{v}})^2}}{2p}\tilde{\mathbf{K}}} e^{-\frac{D\delta}{2pr^2}\tilde{\mathbf{\Theta}}} \right)^p + O\left( \left( \delta/p \right)^2 \right), \tag{38}$$

where the approximation error $O\left( \left( \delta/p \right)^2 \right)$ decreases quadratically with the number of substeps $p$ used to discretize the time interval $\delta$. Therefore, the error can be made arbitrarily small by choosing a sufficiently large value of $p$.

In our application, the computational gain arises from the fact that the matrices $\tilde{\mathbf{\Theta}}$ and $\tilde{\mathbf{K}}$, which are fixed and do not depend on the radius or the sequence parameters (e.g., gradient orientation and gradient strength), now appear inside short-time matrix exponential terms. This allows us to diagonalize $\tilde{\mathbf{K}}$ (i.e., $\tilde{\mathbf{K}} = \mathbf{Q}\mathbf{\Omega}\mathbf{Q}^T$, where $\mathbf{\Omega} = diag\left(\lambda_0,\ldots,\lambda_M\right)$ and $\mathbf{Q}^T\mathbf{Q} = \mathbf{I}$) and write

$$e^{\pm i\frac{qr\sqrt{1-(\hat{\mathbf{g}}\cdot\hat{\mathbf{v}})^2}}{2p}\tilde{\mathbf{K}}} = \mathbf{Q} e^{\pm i\frac{qr\sqrt{1-(\hat{\mathbf{g}}\cdot\hat{\mathbf{v}})^2}}{2p}\mathbf{\Omega}} \mathbf{Q}^T. \tag{39}$$

Notably, this removes the need to evaluate a dense matrix exponential; the exponential in the second term of the previous equation is applied to a diagonal matrix and can be computed by simply taking the scalar exponential of each diagonal entry. Moreover, the diagonalization of $\tilde{\mathbf{K}}$ can be precomputed before the signal evaluation.

By substituting Eq. (39) into Eq. (38), we obtain

$$e^{-\frac{D\delta}{r^2}\tilde{\mathbf{\Theta}} \pm i\frac{qr\sqrt{1-(\hat{\mathbf{g}}\cdot\hat{\mathbf{v}})^2}}{2}\tilde{\mathbf{K}}} \approx e^{-\frac{D\delta}{2pr^2}\tilde{\mathbf{\Theta}}} \mathbf{Q} e^{\pm i\frac{qr\sqrt{1-(\hat{\mathbf{g}}\cdot\hat{\mathbf{v}})^2}}{2p}\mathbf{\Omega}} \mathbf{Q}^T \left( \prod_{i=1}^{p-1} e^{-\frac{D\delta}{pr^2}\tilde{\mathbf{\Theta}}} \mathbf{Q} e^{\pm i\frac{qr\sqrt{1-(\hat{\mathbf{g}}\cdot\hat{\mathbf{v}})^2}}{2p}\mathbf{\Omega}} \mathbf{Q}^T \right) e^{-\frac{D\delta}{2pr^2}\tilde{\mathbf{\Theta}}}. \tag{40}$$

Combining Eq. (40) with Eq. (37), the PGSE signal can be efficiently evaluated. Notice that all matrices involved in the resulting expression, except for the orthogonal matrix $\mathbf{Q}$, are diagonal. In addition, the matrix products appearing in the signal expression do not need to be formed explicitly. Since the measured signal corresponds to the first component of the resulting coefficient vector $\mathbf{c}(0) = [1, \quad 0, \quad \cdots]^T$, it is sufficient to apply the successive matrix exponentials directly to the initial vector. As a result, all operations are carried out in a matrix–vector form, which is computationally cheaper than matrix–matrix products.

This approximation was found to provide an accurate and substantially faster surrogate for repeated model evaluations for different gradient orientations. While the exact signal evaluation described in the previous sections was used to generate all results reported in the **Results** section, we include a dedicated subsection to compare the approximate and exact signals as a function of $p$, and to assess the corresponding computational speed-up, in order to evaluate the practical potential of this approach for future data-driven applications.

### 3.7 Spherical mean signal: Angular averaging and computational strategies

In many dMRI applications, it is desirable to work with observables that are invariant to the orientation of the underlying microstructure. In this context, the spherical mean (powder-averaged) dMRI signal—obtained by averaging the signal over all gradient directions (Edén 2003, Lasič *et al* 2014, Kaden *et al* 2016)—provides a rotationally invariant observable that depends only on the intrinsic properties of the confining geometry and the acquisition parameters. In the present work, we therefore use the spherical mean signal to assess the behavior of the proposed model and to design efficient numerical strategies for its evaluation.

All angular averages reported in this work are computed using spherical Voronoi weights associated with the acquisition gradient directions. Specifically, given a set of unit gradient directions $\{\mathbf{g}_i\}$, the Voronoi-weighted spherical mean signal is computed as

$$\bar{E} = \sum_i w_i E(\mathbf{g}_i), \tag{41}$$

where the weights $w_i$ correspond to the areas of the Voronoi cells on the unit sphere associated with each gradient direction and are normalized such that $\sum_i w_i = 1$. This weighted average mitigates the angular sampling bias arising from non-equidistant gradient direction sets and provides a more accurate approximation of the continuous spherical mean than uniform averaging when the acquisition scheme is not perfectly uniform.

In addition to the scheme-averaged signal, we also compute the ideal spherical mean of the proposed model, defined as the continuous angular average over the unit sphere. Owing to the axial symmetry of the cylindrical surface model in Eq. (37), the spherical mean reduces to a one-dimensional integral over $\mu = \cos(\beta) \in [0,1]$:

$$\frac{\bar{E}}{E(q=0)} = \int_0^1 e^{-q^2(\Delta-\delta/3)D\mu^2} \left[ e^{-\frac{D\delta}{r^2}\tilde{\mathbf{\Theta}} + i\frac{qr\sqrt{1-\mu^2}}{2}\tilde{\mathbf{K}}} e^{-\frac{D(\Delta-\delta)}{r^2}\tilde{\mathbf{\Theta}}} e^{-\frac{D\delta}{r^2}\tilde{\mathbf{\Theta}} - i\frac{qr\sqrt{1-\mu^2}}{2}\tilde{\mathbf{K}}} \right]_{0,0} d\mu \, . \tag{42}$$

This integral is evaluated using Gauss–Legendre quadrature, providing a numerically efficient approximation to the continuous spherical mean. It should be distinguished from the Voronoi-weighted scheme average in Eq. (41), which approximates the angular average from a finite set of acquisition directions.

Unless otherwise stated, comparisons involving the analytical and Monte Carlo signals use the same Voronoi-weighted average over the acquisition scheme. The convergence of the Gauss–Legendre quadrature is assessed separately in **Section 4.3.3** against a high-order evaluation of the continuous spherical mean.

### 3.8 Monte Carlo simulations

Monte Carlo Diffusion Simulations (MCDS) were employed to validate the proposed analytical model. We used an MC simulator developed by our group, available at https://github.com/jonhrafe/Robust-Monte-Carlo-Simulations (Rafael-Patino *et al* 2020, Villarreal-Haro *et al* 2023). This tool has been validated against analytical models across multiple geometries, including impermeable planes, cylinders, spheres, and more recently in cylindrical surfaces (Canales-Rodríguez *et al* 2025).

We compared the analytical dMRI signal to those generated by the MC simulations for identical cylindrical surfaces. Specifically, dMRI signals were generated from 50 independent cylinders with radii uniformly spaced from 0.1 μm to 5.0 μm in increments of 0.1 μm following the protocol described in (Canales-Rodríguez *et al* 2025). The diffusion process simulation was carried out using $N_p$ = 75×10$^3$ particles uniformly distributed on each cylindrical surface and $N_t$ = 15×10$^3$ diffusion steps per particle. The axial myelin water diffusivity was set to $D$ = 0.8 μm$^2$/ms, as this value produced the largest discrepancies in our previous model (Canales-Rodríguez *et al* 2025).

The dMRI data were generated using a PGSE sequence with trapezoidal diffusion gradients employing a gradient strength of $G$ = 500 mT/m and a slew rate of $SR$ = 500 T/m/s, which are compatible with the specifications of a Connectome 2.0 3T scanner (Huang *et al* 2021) and small animal 7T and 9.4T BioSpec MRI scanners. The protocol included 90° and 180° pulse durations of 2 ms and 4 ms, respectively. The data were generated for six $b$-values = [1, 2, 3, 4, 5, 6] ms/μm$^2$, which were selected using the shortest possible echo time (TE) for each case, while keeping the TE smaller than 20 ms and maintaining maximum $G$ and $SR$, following the implementation described in (Mueller *et al* 2019, Tax *et al* 2019). For each $b$-value, dMRI signals were generated for 92 gradient orientations uniformly distributed on the unit sphere, along with the signal for $b$ = 0. Table I shows the experimental parameters.

The subsequent analyses focused on the spherical mean signal normalized by the $b$ = 0 signal. The same Voronoi weights are consistently applied to analytical model evaluations and Monte Carlo simulations.

**Table I**. Experimental parameters for Monte Carlo simulations using a PGSE sequence.

| $b$-value (ms/μm$^2$) | $\Delta$ (ms) | $\delta$ (ms) | TE (ms) |
|---|---|---|---|
| 1.0 | 7.72 | 2.88 | 13.43 |
| 2.0 | 8.72 | 3.89 | 15.44 |
| 3.0 | 9.45 | 4.62 | 16.90 |
| 4.0 | 10.03 | 5.20 | 18.06 |
| 5.0 | 10.53 | 5.70 | 19.06 |
| 6.0 | 10.97 | 6.14 | 19.94 |

## 4 Results

In this section, we first illustrate characteristic features of the diffusion signal predicted by the proposed cylindrical surface model and validate the exact analytical formulation against Monte Carlo simulations using scheme-averaged spherical mean signals computed with Voronoi weights. We then investigate the numerical accuracy and computational efficiency of the spectral formulation by analyzing its convergence with respect to the truncation order of the spectral basis and by introducing two complementary acceleration strategies based on operator splitting and angular quadrature.

### 4.1 *b-value*–dependent spherical mean signal profile

Figure 1 illustrates the theoretical spherical mean dMRI signal arising from diffusion confined to a cylindrical surface, as predicted by the general analytical model introduced in Eq. (37). The signal was computed for a PGSE sequence with $b$ *-values* ranging from 0 to 40 ms/µm², and three representative cylinder radii: 1.0 µm, 2.0 µm, and 3.0 µm. Additionally, we plot the spherical mean dMRI signal from the recently proposed spherical-mean Gaussian phase approximation (GPA) model for this geometry (Lee *et al* 2026).

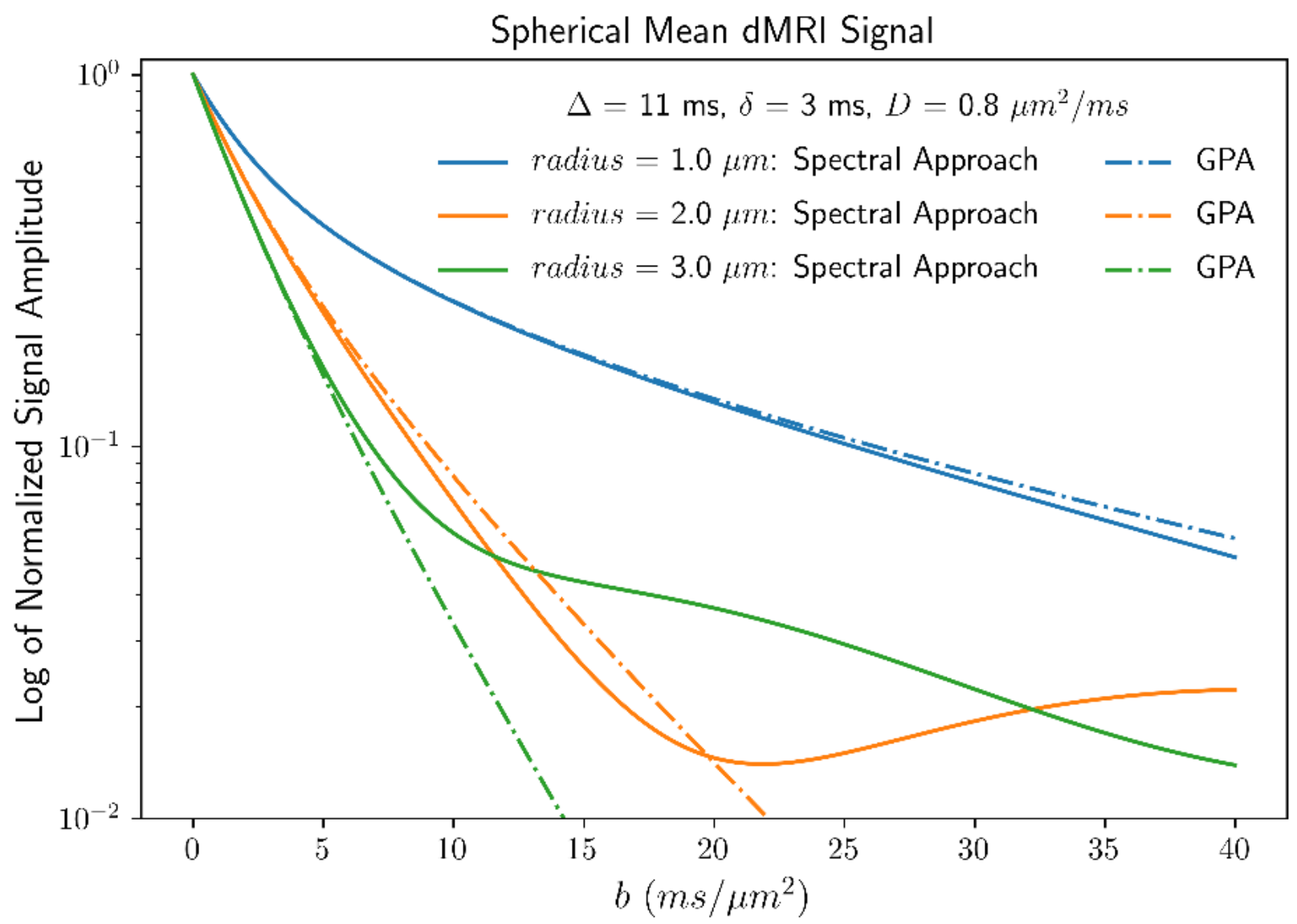

**Figure 1.** *Signal attenuation for diffusion confined to cylindrical surfaces.* The spherical mean dMRI signal was generated using the analytical model introduced in Eq. (37) for a PGSE sequence with $\Delta$ = 11 ms and $\delta$ = 3 ms. The axial diffusivity was set to $D$ = 0.8 µm²/ms. The signal attenuation is shown on a logarithmic scale for three cylinder radii: 1.0 µm (blue), 2.0 µm (orange), and 3.0 µm (green). The analytical expression was evaluated using $M$ =50 spectral modes (continuous lines). Additionally, we show the spherical mean signal resulting from the Gaussian Phase Approximation (GPA) model (Lee *et al* 2026) (dashed lines). Note that the large *b*-values used in this figure are intended to illustrate diffraction-like patterns that are not necessarily accessible with standard clinical protocols.

For small radii (e.g., $r$ =1.0 µm), the GPA closely follows the exact analytical signal over a wide range of $b$ -values, with only minor deviations appearing at very high diffusion weightings ( $b \geq$ 20ms/µm$^2$). As the radius increases, however, noticeable differences between the two models emerge at lower $b$ -values. For $r$ =2.0 µm and $r$ =3.0 µm, the GPA progressively deviates from the exact solution for $b \geq 5$ ms/µm$^2$, with the discrepancy becoming more pronounced for larger $b$ -values. At very high diffusion weightings, the exact solution exhibits oscillatory features commonly referred to as diffraction patterns, which arise from the angular confinement of diffusion and the resulting interference among discrete Laplacian eigenmodes on the cylindrical surface. These features are not captured by the GPA, highlighting the importance of the exact analytical formulation when describing the signal in regimes where restricted diffusion effects become dominant.

### 4.2 Validation against Monte Carlo simulations

To validate the accuracy of the proposed analytical model, we compared its predictions against MC-generated signals. Analytical signals were computed using $M$ = 50 modes.

Figure 2 shows the spherical mean dMRI signal as a function of cylinder radius for an axial diffusivity of $D$ $D$= 0.8 μm$^2$/ms, and six $b$-*values* ranging from 1.0 to 6.0 ms/μm$^2$. The analytical spectral model (continuous lines) exhibits excellent agreement with the MC simulations (discrete points) across the entire range of radii and $b$-values considered. The GPA predictions also closely reproduce the MC signals for small cylinder radii, but deviations progressively emerge as the radius increases. These deviations become more pronounced at higher diffusion weightings, for which the GPA increasingly underestimates the signal at larger radii. In contrast, the spectral formulation maintains excellent agreement with the MC simulations over the full range investigated. This confirms the correctness of the spectral formulation and the exact treatment of finite gradient pulse effects.

As expected, increasing the $b$-*value* leads to stronger signal attenuation and enhanced sensitivity to the cylinder radius. For the acquisition protocol considered here, the signal displays limited sensitivity to radii smaller than approximately 0.5 μm and larger than approximately 4.0 μm, indicating a finite radius sensitivity window determined by the gradient strength and diffusion times of the PGSE protocol.

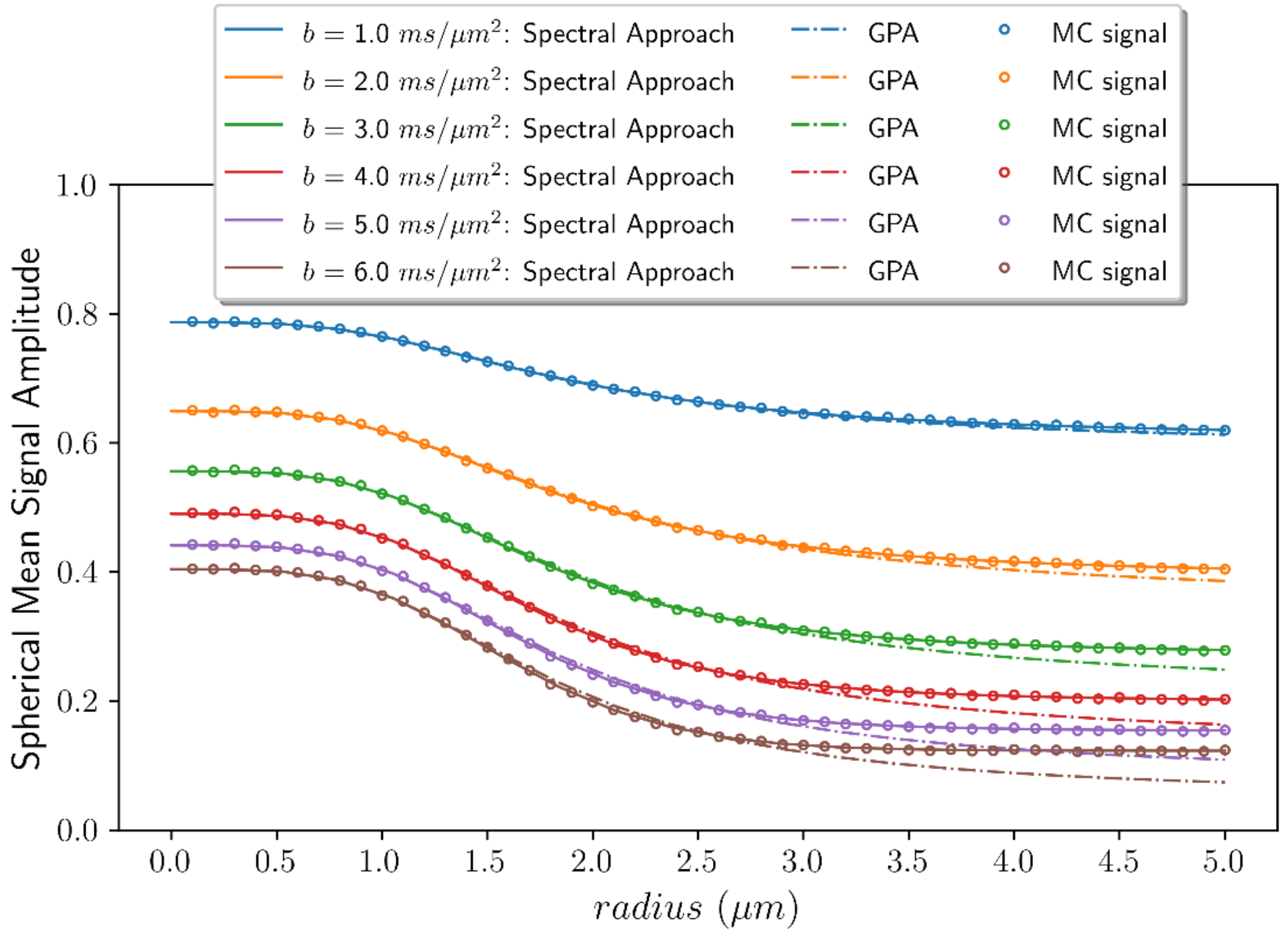

**Figure 2.** *Spherical mean dMRI signals and Monte Carlo simulations.* Spherical mean dMRI signals are shown as a function of cylinder radius, computed using the analytical spectral model (continuous lines), Monte Carlo (MC) simulations (dots), and the Gaussian Phase Approximation (GPA) model (Lee *et al* 2026) (dashed lines). Signals are evaluated for $b$-values of [1.0, 2.0, 3.0, 4.0, 5.0, 6.0] ms/μm$^2$, using the PGSE sequence parameters listed in Table I and an axial diffusivity of $D$=0.8 μm$^2$/ms. Analytical signals and MC simulations were generated for 50 discrete radii spanning 0–5. μm.

### 4.3 Numerical analyses

To characterize the numerical accuracy and computational performance of the proposed analytical formulation, we investigate three independent sources of approximation error: (i) the truncation order of the spectral expansion, (ii) the number of substeps used in the Strang splitting approximation, and (iii) the number of quadrature nodes employed to evaluate the spherical mean.

In all cases, we quantify the associated accuracy-runtime trade-offs in order to assess the practical suitability of the proposed methods for repeated signal evaluations, as required in parameter estimation and fitting applications.

All experiments were performed using the same PGSE parameters $\Delta$ = 9.45 ms and $\delta$ = 4.61 ms, the same set of 92 diffusion-encoding directions used in the Monte Carlo simulations, and the same three diffusion weightings $b$ = [3, 6, 20] ms/µm². Unless otherwise stated, the spherical mean is computed using the Voronoi-weighted average over the gradient orientations. The analysis was carried out for $N_r$=50 independent cylinders with radii uniformly distributed between 0.1 µm and 5.0 µm.

For all benchmarks, execution was restricted to a single CPU core and a single thread in order to enable a fair comparison of the intrinsic computational cost, independent of hardware-specific parallelization or multi-threading effects. All computations were performed on a laptop equipped with an Intel Core i7-7700HQ CPU (8 cores, 2.80 GHz) running Linux.

These benchmarks are intended to characterize the accuracy–runtime trade-offs for a representative PGSE protocol rather than to define universal numerical settings. The required truncation order, number of Strang substeps, and number of quadrature nodes should be reassessed when the acquisition or model-parameter ranges are substantially changed.

#### 4.3.1 Convergence of the analytical signal with respect to the truncation order

We first investigated the convergence of the analytical signal with respect to the truncation order $M$ of the spectral expansion. The Voronoi-weighted spherical mean computed using $M$ = 50 modes was taken as reference, denoted by $\bar{E}_{ref}$. Voronoi-weighted spherical mean signals $\bar{E}_M$ computed using lower truncation orders $M$ = [3, 5, 10, 20] were compared against this reference.

For each truncation order and each $b$-value, the total signal discrepancy over all $N_r$=50 radii was quantified using the mean relative absolute error (*MRAE*), defined as

$$MRAE_M = \frac{1}{N_r}\sum_{i=1}^{N_r}\frac{\left|\bar{E}_M\left(r_i\right)-\bar{E}_{ref}\left(r_i\right)\right|}{\bar{E}_{ref}\left(r_i\right)}\cdot 100\%. \tag{43}$$

In addition, we report the computation time required to generate the Voronoi-weighted spherical mean signal for all gradient directions per radius, together with the corresponding acceleration factor relative to the reference computation using $M$ =50 modes. For comparison, we also include the results obtained from the spherical-mean GPA model (Lee *et al* 2026).The resulting accuracy and performance metrics are summarized in Table II.

The results demonstrate very rapid convergence of the analytical signal with respect to the truncation order. In particular, $M$ =10 already yields results that are numerically indistinguishable from those obtained with substantially higher truncation orders, while even $M$ =5 provides accurate approximations under the present acquisition conditions. This confirms that only a limited number of angular modes contribute significantly to the signal, since the quadratic growth of the Laplace eigenvalues strongly suppresses higher-order contributions.

For comparison, the spherical-mean GPA model provides the shortest computation time, yielding an acceleration factor of approximately $4.8\times10^3$ relative to the reference calculation. However, this gain in speed comes at the cost of substantially larger relative absolute errors, exceeding those obtained even with very low truncation orders (e.g., $M$ =3).

**Table II.** Mean relative absolute error (*MRAE*), computation time and acceleration factor as a function of the truncation order $M$ of the spectral expansion. The *MRAE* and acceleration factor are computed with respect to the Voronoi-weighted spherical mean signal obtained with $M$ =50 modes. Reported computation times correspond to the average time required to generate the Voronoi-weighted spherical mean signal over the 92 diffusion-encoding directions for a single cylinder radius and $b$ -value. The average was computed by dividing the total runtime for all 50 radii and three diffusion weightings by 50×3. Results are reported for three diffusion weightings: $b$ = [3, 6, 20] ms/µm². Results from the spherical mean Gaussian phase approximation (GPA) model are also included.

| Truncation order | *MRAE* (100%) | | | Computation time (*milliseconds*) | Acceleration factor (×) |
|---|---|---|---|---|---|
| | $b$ =3.0 ms/µm² | $b$ =6.0 ms/µm² | $b$ =20 ms/µm² | | |
| $M$ = 3 | $4.1\times10^{-3}$ | $6.2\times10^{-2}$ | 4.5 | 9.8 ms | 11.3 |
| $M$ = 5 | $2.8\times10^{-7}$ | $2.4\times10^{-5}$ | $5.1\times10^{-2}$ | 10.2 ms | 10.9 |
| $M$ = 10 | $3.8\times10^{-11}$ | $1.8\times10^{-11}$ | $2.5\times10^{-10}$ | 12.5 ms | 8.9 |
| $M$ = 20 | $3.4\times10^{-11}$ | $2.4\times10^{-11}$ | $3.2\times10^{-11}$ | 18.5 ms | 6.0 |
| $M$ = 50 | *Reference* | *Reference* | *Reference* | 111.0 ms | *Reference* |
| GPA | 3.05 | 14.6 | 61.0 | 0.023 ms | $4.8\times10^3$ |

### 4.3.2 Convergence of the approximated signal with respect to the Strang splitting

Next, we assessed the convergence properties of the approximated analytical signal obtained using the Strang splitting approach as a function of the number of substeps $p$.

As reference, we used the Voronoi-weighted spherical mean signal generated with the exact analytical model using M=10 modes, which was selected based on the convergence analysis of the previous subsection. Approximated signals computed using $p$ =[5, 10, 20, 40, 80] substeps were compared against this reference.

The analysis was performed for the same set of 50 cylinder radii and the same three $b$-values. For each value of $p$, the overall discrepancy with respect to the reference signal was quantified using the *MRAE*. In addition, we report the corresponding computation times and acceleration factors relative to the exact analytical evaluation. The results are reported in Table III.

**Table III.** Mean relative absolute error (*MRAE*), computation time and acceleration factor as a function of the number of Strang splitting substeps $p$. The *MRAE* and acceleration factor are computed with respect to the exact analytical signal generated with $M$ =10 modes. Reported computation times correspond to the average time required to generate the Voronoi-weighted spherical mean signal over the 92 diffusion-encoding directions for a single cylinder radius and $b$-value. The average was computed by dividing the total runtime for all 50 radii and three diffusion weightings by 50×3. Results are reported for three diffusion weightings: $b$ = [3, 6, 20] ms/µm².

| Splitting substeps | *MRAE* (100%) | | | Computation time (*milliseconds*) | Acceleration factor (×) |
|---|---|---|---|---|---|
| | $b$ =3.0 ms/µm² | $b$ =6.0 ms/µm² | $b$ =20 ms/µm² | | |
| $p$ = 5 | 0.55 | 1.01 | 2.0 | 0.6 ms | 20.8 |
| $p$ = 10 | 0.14 | 0.26 | 0.53 | 0.8 ms | 15.6 |
| $p$ = 20 | 0.036 | 0.065 | 0.135 | 1.3 ms | 9.6 |
| $p$ = 40 | 0.009 | 0.016 | 0.034 | 2.2 ms | 5.7 |
| $p$ = 80 | 0.002 | 0.004 | 0.009 | 4.1 ms | 3.0 |

The results in Table III show that the Strang splitting approximation achieves high accuracy with a relatively moderate number of substeps, while substantially reducing the computational cost. For example, for $p$ =20 the *MRAE* is below 0.2% across all diffusion weightings, while the computation is approximately one order of magnitude faster than the exact evaluation. For $p$ =40, the approximated signals are visually indistinguishable from the exact signals (results not shown).

A small number of substeps leads to noticeable approximation errors, particularly at higher diffusion weightings, highlighting the need for a minimal number of substeps when operating in high-$b$ regimes.

### 4.3.3 Convergence of the spherical mean with respect to quadrature nodes

Since the spherical mean constitutes a central observable in many applications, we investigated the convergence properties of its numerical evaluation using Gauss–Legendre quadrature.

All signals in this experiment were generated using the exact analytical model with $M$ =10 modes. To isolate the quadrature error from effects associated with the finite angular sampling of the acquisition scheme, the reference continuous spherical mean was computed using Gauss–Legendre quadrature with $n_q$ =100 nodes. Lower-order quadratures with $n_q$ =[5,6,7,8,20] nodes were then compared against this reference.

For each value of $n_q$ and each $b$-value, the discrepancy with respect to the $n_q$ =100 reference was quantified using the MRAE over the same 50 cylinder radii considered in the previous sections. The corresponding computation times and acceleration factors are reported in Table IV.

**Table IV.** Mean relative absolute error (*MRAE*), computation time, and acceleration factor as a function of the number of quadrature nodes $n_q$ used in the Gauss–Legendre integration. All signals were generated using the exact analytical model with $M$ =10 modes. The *MRAE* is computed with respect to the continuous spherical mean evaluated using Gauss–Legendre quadrature with $n_q$ =100 nodes. Reported computation times correspond to the average time required to evaluate the spherical mean using the indicated number of Gauss–Legendre nodes for a single cylinder radius and $b$-value. The average was computed by dividing the total runtime for all 50 radii and three diffusion weightings by 50×3. To maintain consistency with Tables II and III, the acceleration factor is computed relative to the exact analytical scheme-averaged signal evaluated over the 92 diffusion-encoding directions using Voronoi weighting. Results are reported for three diffusion weightings: $b$ =[3,6,20] ms/μm².

| Quadrature nodes | *MRAE* (100%) | | | Computation time (*milliseconds*) | Acceleration factor (×) |
|---|---|---|---|---|---|
| | $b$ =3.0 ms/μm² | $b$ =6.0 ms/μm² | $b$ =20 ms/μm² | | |
| $n_q$ = 5 | $4.93\times10^{-5}$ | $1.37\times10^{-4}$ | $6.23\times10^{-2}$ | 0.7 ms | 17.9 |
| $n_q$ = 6 | $1.02\times10^{-6}$ | $3.10\times10^{-5}$ | $2.05\times10^{-2}$ | 0.9 ms | 13.9 |
| $n_q$ = 7 | $1.60\times10^{-8}$ | $2.39\times10^{-6}$ | $4.42\times10^{-3}$ | 1.0 ms | 12.5 |
| $n_q$ = 8 | $1.81\times10^{-10}$ | $1.09\times10^{-7}$ | $4.05\times10^{-4}$ | 1.2 ms | 10.4 |
| $n_q$ = 20 | $1.80\times10^{-12}$ | $2.09\times10^{-12}$ | $3.21\times10^{-12}$ | 2.8 ms | 4.5 |

The results demonstrate rapid convergence of the Gauss–Legendre quadrature. With only $n_q$ =5 nodes, the *MRAE* is $4.93\times10^{-5}$, $1.37\times10^{-4}$, and $6.23\times10^{-2}$ for $b$ =3, 6, and 20 ms/μm², respectively. Increasing the number of nodes further reduces the quadrature error. At $n_q$ =20, the differences with respect to the $n_q$ =100 reference are of the order of $10^{-12}$ for all three diffusion weightings. For the protocol considered here, this indicates that the spherical mean can be accurately evaluated using substantially fewer quadrature nodes than diffusion-encoding directions. This enables a significant reduction in the computational cost of spherical mean evaluation, without requiring the explicit generation of the signal for each experimental gradient direction.

To separately quantify the effect of finite angular sampling, we also compared the Voronoi-weighted average over the 92 acquisition directions with the converged $n_q$ =100 spherical mean. The corresponding MRAE values were 0.026%, 0.078%, and 0.653% for $b$ =3, 6, and 20 ms/μm², respectively. These residual differences reflect the finite angular sampling of the acquisition scheme.

After evaluating the individual acceleration strategies in this and previous sections, we finally assess their combined impact on the computation of the spherical mean signal. To this end, we performed a representative experiment in which the spherical mean analytical signal was generated using a truncation order of $M$ =10, evaluated through the Strang splitting approach with $p$ =10 steps and $n_q$ =5 Gauss–Legendre quadrature nodes. Using the Voronoi-weighted exact signal as reference, the resulting *MRAE* values were [0.17, 0.34, 1.11] for the three considered diffusion weightings ($b$ = [3, 6, 20] ms/µm²), respectively. The average computation time was 0.031 ms per cylinder radius and $b$-value. The implementation was further accelerated by compiling the numerical routines using the just-in-time Numba compiler (Lam *et al* 2015).

Comparing these results with those obtained from the GPA model reported in Table II for the same acquisition parameters shows that the computational cost of the accelerated analytical formulation is of the same order of magnitude as that of the GPA (i.e., 0.031 ms vs 0.023 ms), while yielding substantially smaller approximation errors.

## 5 Discussion

The main result of this work is the exact analytical solution of the Bloch–Torrey equation for diffusion confined to a cylindrical surface under finite PGSE gradients, yielding a closed-form expression for the corresponding dMRI signal valid for arbitrary gradient durations and separations. Notably, this solution does not rely on assumptions on the propagator, spin phase distribution, or finite-pulse effects. Related spectral approaches have been reported in the literature for restricted diffusion inside cylinders and spheres, and between parallel planes with reflecting boundaries (Grebenkov 2008b, Caprihan *et al* 1996, Sukstanskii and Yablonskiy 2002, Callaghan 1997, Barzykin 1999).

The only approximation involved in the practical evaluation of the signal expression is numerical, due to the truncation of the spectral basis. This truncation is well controlled, as the eigenvalues of the Laplace operator on the cylindrical surface grow quadratically with the mode index, leading to rapid convergence of the series. As a result, the analytical signal can be evaluated with high accuracy using a relatively small number of spectral modes (e.g., $M$ =10; see Table II). The proposed analytical signal expression was validated against Monte Carlo diffusion simulations over a wide range of cylinder radii and PGSE parameters (see Figure 2).

For the geometry under study, the diffusion process is constrained along the (periodic) angular coordinate while remaining unbounded along the cylinder's main axis. This leads to a natural decoupling of the diffusive motion into perpendicular and parallel components, reflected in the factorized structure of the signal. We significantly reduced the dimensionality of the matrices involved in the signal evaluation by exploiting the symmetry of the cylindrical surface and introducing a reduced real spectral basis, thus achieving computational gains without compromising accuracy.

Beyond the exact analytical formulation, an additional contribution of this work is the introduction of efficient numerical strategies that enable fast and repeated evaluations of the proposed signal model. In particular, we introduced a second-order Strang splitting approximation for the matrix exponential operators arising in the PGSE signal expression, which allows the time evolution to be decomposed into a sequence of simpler operators that can be efficiently applied through a precomputed eigendecomposition. This approach avoids repeated evaluations of dense matrix exponentials and enables the signal to be computed by a sequence of matrix–vector operations, yielding substantial computational savings while preserving a controllable approximation error (see Table III). In addition, we proposed an efficient evaluation of the spherical mean (powder-averaged) signal based on Gauss–Legendre quadrature, exploiting the axial symmetry of the cylindrical surface model to reduce the angular integration to a one-dimensional quadrature problem (see Table IV). Together, these developments provide a practical framework for accelerating both the evaluation of the directional signal and the computation of rotationally invariant observables, which are of central interest in many dMRI applications and in parameter estimation problems requiring repeated model evaluations.

The numerical analyses presented in this work further clarify the accuracy–efficiency trade-offs associated with these approximations. Although the results show that relatively small truncation orders of the spectral expansion and a moderate number of Strang splitting substeps are sufficient to achieve high accuracy for the PGSE protocol considered here, these numerical parameters should not be adopted indiscriminately. In practical applications, the required truncation order depends on the acquisition parameters and on the range of cylinder radii of interest, and a dedicated convergence analysis should therefore be carried out for each specific application. Likewise, the convergence of the Strang splitting approximation depends on the temporal structure and strength of the diffusion-encoding gradients, so that an application-specific assessment is needed to identify an appropriate number of splitting substeps that balances accuracy and computational efficiency. The number of Gauss–Legendre nodes required to achieve a prescribed accuracy similarly depends on the angular complexity of the signal and should therefore be verified for the acquisition and model-parameter regime of interest. Finally, the Voronoi-weighted scheme average and the Gauss–Legendre evaluation correspond to two different approximations of angular averaging. The Voronoi-based approach estimates the spherical mean from the finite set of acquired diffusion-encoding directions, whereas Gauss–Legendre quadrature approximates the continuous, rotationally invariant spherical mean of the model. The convergence analysis in Table IV shows that the Gauss–Legendre quadrature rapidly converges to the continuous spherical mean. The residual difference between this converged result and the Voronoi-weighted average therefore reflects the finite angular sampling of the acquisition scheme rather than incomplete convergence of the Gauss–Legendre quadrature.

In a previous study, we derived an exact analytical expression for diffusion constrained to cylindrical surfaces in the narrow-pulse limit and proposed an approximate extension to finite gradient durations (Canales-Rodríguez *et al* 2025). While that formulation captured the main features of surface-confined diffusion and showed good agreement with Monte Carlo simulations in various regimes, discrepancies emerged for large diffusivities and high diffusion weightings, where finite-pulse and non-Gaussian effects become non-negligible. Recently, a new extension for wide pulses based on the Gaussian phase approximation was introduced by (Lee *et al* 2026). As shown in Figure 2, this approximation remains highly accurate for relatively small cylinder radii, but progressively deviates from the Monte Carlo reference as the radius increases, particularly at higher diffusion weightings. The present work overcomes previous limitations by providing an exact treatment of finite gradient pulse durations, thereby completing the theoretical description presented in previous works.

From a broader theoretical perspective, a general spectral approach was developed to describe restricted diffusion in circular layers of arbitrary thickness under time-dependent gradient fields (Grebenkov 2008a). Within such a framework, diffusion confined to a cylindrical surface can be formally interpreted as a limiting case corresponding to a vanishing radial layer thickness. Therefore, our work could be rederived following the theoretical treatment presented in (Grebenkov 2008a). However, rather than relying on this implicit limit, the present work follows a direct derivation tailored to the surface geometry. This strategy enables the explicit construction of the spectral geometry-specific matrices governing the PGSE signal and yields a compact analytical expression that is particularly suited for practical signal evaluation.

The present formulation relies on a number of idealizations that define its domain of applicability. The cylinder is assumed to be infinitely long, thereby neglecting end effects. This assumption can, however, be relaxed without modifying the circumferential component of the model. For a cylindrical surface of finite axial length $L$, the angular and axial motions remain separable, and the free-diffusion factor in Eq. (36) can be replaced by the signal for one-dimensional restricted diffusion on the interval $0 \le z \le L$ with reflecting boundary conditions at both ends. In addition, the surface is treated as perfectly smooth and impermeable, and no exchange with other compartments is considered. Moreover, relaxation effects and surface relaxivity are not explicitly included in the model. While these assumptions are appropriate for isolating the effect of surface confinement on the diffusion signal, they should be kept in mind when considering applications to biological tissues or more complex geometries. The analytical framework developed here can be extended in several directions. Possible extensions include the incorporation of distributions of (concentric) cylinder radii, alternative gradient waveforms, and additional physical effects such as relaxation or exchange.

The presented model can be extended to arbitrary waveforms following the approach described in (Grebenkov 2009, Caprihan *et al* 1996, Callaghan 1997). That is, by dividing the time interval of the waveform into small subintervals and approximating the waveform by a piecewise constant function in each subinterval. The optimal number of subintervals depends on the smoothness of the waveform. For example, a PGSE experiment with trapezoidal gradients would require at least seven subintervals to represent the full waveform, including the rising and falling ramps of the two diffusion gradients, the two plateau intervals at constant gradient amplitude, and the interval in which the gradient is zero. In practice, a finer subdivision of the ramp intervals may be required depending on the slew rate and on the desired accuracy of the waveform approximation. The resulting signal would involve the multiplication of the matrix exponentials corresponding to each of the subintervals in which the waveform is assumed to be constant. As a result, although the practical implementation is straightforward, the computation time grows proportionally with the number of subintervals.

For this reason, we focused on rectangular PGSE gradients, as they involve three distinct subintervals and allow an exact analytical treatment. In modern scanners with high slew rates, the ramp durations of trapezoidal gradients are short compared to the pulse duration, so their impact on the resulting diffusion signal is relatively small. As a result, trapezoidal gradients produce signals that are very well approximated by those from rectangular gradients with the same nominal timing parameters $(\Delta, \delta)$. Consistent with this, the signals generated by the proposed analytical model were in excellent agreement with those obtained from Monte Carlo simulations using trapezoidal gradients (see Table I and Figure 2).

While the cylindrical surface model is motivated by applications to myelin water diffusion (Canales-Rodríguez *et al* 2025, Avram *et al* 2010, Brusini *et al* 2019, Tax *et al* 2019, Andrews *et al* 2006, Peled *et al* 1999, Lee *et al* 2026), this work focuses on the theoretical formulation of the dMRI signal model, establishing a rigorous foundation for future experimental and data-driven studies. Closed-form analytical solutions of the Bloch–Torrey equation under finite gradient pulses are available only for a limited number of geometries. The present result extends this class to diffusion confined to cylindrical surfaces.

**Data availability statement**

The datasets and source code supporting the findings of this study, including implementations of the exact analytical model and the accelerated numerical strategies presented in this work, are publicly available at the following repository: https://github.com/ejcanalesr/exact-PGSE-dMRI-signal-cylindrical-surface.

## Author contributions

EC-R: Conceptualization, Investigation, Methodology, Project administration, Data curation, Formal Analysis, Resources, Software, Supervision, Validation, Visualization, Writing–original draft, Writing–review and editing. CT: Investigation, Software, Validation, Visualization, Writing–review and editing. JM-G: Investigation, Resources, Writing–review and editing. DJ: Investigation, Resources, Writing–review and editing. J-PT: Investigation, Resources, Writing–review and editing. JR-P: Investigation, Methodology, Data curation, Formal Analysis, Resources, Software, Validation, Visualization, Writing–original draft, Writing–review and editing.

## Acknowledgments

E.J.C.R. acknowledges financial support from the “Ramón y Cajal” Excellence Fellowship (Grant RYC2023-042763-I), funded by the Ministry of Science, Innovation and Universities (MICIU) and the Spanish State Research Agency (AEI, 10.13039/50110001103), and co-financed by the European Social Fund Plus (ESF+).

## Conflict of interest

The authors declare that the research was conducted without any commercial or financial relationships that could be construed as a potential conflict of interest.

## Generative AI statement

During the preparation of this work the authors used ChatGPT in order to assist in identifying grammatical errors and typos in this manuscript. After using this tool/service, the authors reviewed and edited the content as needed and take full responsibility for the content of the published article.

## Appendices

### Appendix A: Computing matrix B

The integral in Eq. (17) is solved as:

$$\begin{aligned} B_{mn} &= \left(\frac{1}{2\pi}\right) r \int_0^{2\pi} \cos(\theta) e^{in\theta} e^{-im\theta} d\theta \\ &= \left(\frac{1}{2\pi}\right) r \int_0^{2\pi} \cos(\theta) e^{i(n-m)\theta} d\theta \\ & \left(\frac{1}{2\pi}\right) \frac{r}{2} \int_0^{2\pi} \left(e^{i\theta} + e^{-i\theta}\right) e^{i(n-m)\theta} d\theta \\ &= \left(\frac{1}{2\pi}\right) \frac{r}{2} \left( \int_0^{2\pi} e^{i(n-(m-1))\theta} d\theta + \int_0^{2\pi} e^{i(n-(m+1))\theta} d\theta \right), \end{aligned} \tag{44}$$

where we used Euler´s formula. Without loss of generality, here we assumed that the diffusion gradient is applied along the x-axis (assuming the cylinder’s axis is parallel to the z-axis). Therefore, the angle between the gradient and the position vector is $\theta$. Note this assumption does not affect the dMRI signal $E_{\perp}$, since it is constant for any gradient orientation lying on the plane perpendicular to the cylinder’s axis.

The above integrals can be evaluated by using the identity:

$$\int_0^{2\pi} e^{i(n-m)\theta} d\theta = 2\pi\delta_{n,m} \quad . \tag{45}$$

Thus,

$$B_{mn} = \frac{r}{2}\left(\delta_{n,m-1} + \delta_{n,m+1}\right). \tag{46}$$

**Appendix B: Symmetry of the spectral coefficients and construction of the compact basis**

Starting from Eq. (6), here we show that the coefficients associated with opposite modes satisfy $c_m(t) = c_{-m}(t)$. Using the expressions obtained in **Section 3.1** for the eigenvalues and for the gradient-coupling matrix (see Eq. (17) or Eq. (46)), the evolution equation for a coefficient of order $m$ can be written as

$$\begin{aligned} \frac{\partial c_m(t)}{\partial t} &= -Dc_m(t)\lambda_m - i\gamma G(t)\frac{r}{2}\sum_n c_n(t)\left(\delta_{n,m-1} + \delta_{n,m+1}\right), \\ &= -Dc_m(t)\lambda_m - i\gamma G(t)\frac{r}{2}\left(c_{m-1}(t) + c_{m+1}(t)\right). \end{aligned} \tag{47}$$

The corresponding equation for the coefficient of order $-m$ is

$$\frac{\partial c_{-m}(t)}{\partial t} = -Dc_{-m}(t)\lambda_{-m} - i\gamma G(t)\frac{r}{2}\left(c_{-(m-1)}(t) + c_{-(m+1)}(t)\right). \tag{48}$$

It is important to note that as $\lambda_m = \lambda_{-m} = \frac{m^2}{r^2}$, both equations have the same functional form and therefore are invariant under the transformation $m \to -m$. Furthermore, for the spatially uniform initial magnetization considered in this work, only the ground mode is initially populated, so that

$$c_m(0) = c_{-m}(0) = 0, \quad m > 0 \,. \tag{49}$$

This follows from the expansion in Eq. (4) for $t = 0$, together with the observation that the only spatially constant eigenfunction is the ground state $\phi_0(\mathbf{r})$.

The coefficients $c_m(t)$ and $c_{-m}(t)$ thus satisfy identical systems of differential equations with identical initial conditions. Then, by uniqueness of the solution,

$$c_m(t) = c_{-m}(t), \quad m > 0, \tag{50}$$

at all times during the diffusion-encoding sequence. Since the mode index $m$ is arbitrary, the same result may equivalently be written as $c_n(t) = c_{-n}(t)$, using the notation adopted in **Section 3** for the cylindrical modes.

**Appendix C: Proving the relationship in Eq. (27)**

The transformation matrix in Eq. (25) satisfies the following properties: $\mathbf{T}^T\mathbf{T} = \mathbf{I}$ is the identity and

$$
\mathbf{T}\mathbf{T}^T = \begin{bmatrix} 1/2 & \cdots & 0 & 0 & 0 & \cdots & 1/2 \\ \vdots & \ddots & & \vdots & & \iddots & \vdots \\ 0 & & 1/2 & 0 & 1/2 & & 0 \\ 0 & \cdots & 0 & 1 & 0 & \cdots & 0 \\ 0 & & 1/2 & 0 & 1/2 & & 0 \\ \vdots & \iddots & & \vdots & & \ddots & \vdots \\ 1/2 & \cdots & 0 & 0 & 0 & \cdots & 1/2 \end{bmatrix}. \tag{51}
$$

For the matrices appearing in Eqs. (26), the following identities can be verified directly from the explicit forms of $\mathbf{T}$, $\boldsymbol{\Lambda}$, and $\mathbf{B}$: $\boldsymbol{\Lambda}\mathbf{T} = \mathbf{T}\left(\mathbf{T}^T\boldsymbol{\Lambda}\mathbf{T}\right)$ and $\mathbf{B}\mathbf{T} = \mathbf{T}\left(\mathbf{T}^T\mathbf{B}\mathbf{T}\right)$. Therefore, for any linear combination of $\boldsymbol{\Lambda}$ and $\mathbf{B}$, $\mathbf{A} = a\boldsymbol{\Lambda} + b\mathbf{B}$, it follows that: $\mathbf{A}\mathbf{T} = \mathbf{T}\left(\mathbf{T}^T\mathbf{A}\mathbf{T}\right)$.

The relationship we want to simplify is $\mathbf{T}^T e^{-\mathbf{A}_3} e^{-\mathbf{A}_2} e^{-\mathbf{A}_1}\mathbf{T}$. For this, let´s first write the matrix exponential in Taylor series:

$$
e^{-\mathbf{A}} = \sum_{k=0}^{\infty} \frac{(-1)^k}{k!}\mathbf{A}^k = \mathbf{I} - \mathbf{A} + \frac{1}{2!}\mathbf{A}^2 - \frac{1}{3!}\mathbf{A}^3 + \ldots. \tag{52}
$$

After substituting this relationship into the term $e^{-\mathbf{A}_1}$ we get

$$
\begin{aligned}
\mathbf{T}^T e^{-\mathbf{A}_3} e^{-\mathbf{A}_2} e^{-\mathbf{A}_1}\mathbf{T} &= \mathbf{T}^T e^{-\mathbf{A}_3} e^{-\mathbf{A}_2} \left( \sum_{k=0}^{\infty} \frac{(-1)^k}{k!}\mathbf{A}_1^k\mathbf{T} \right), \\
&= \mathbf{T}^T e^{-\mathbf{A}_3} e^{-\mathbf{A}_2}\mathbf{T} \left( \sum_{k=0}^{\infty} \frac{(-1)^k}{k!}\left(\mathbf{T}^T\mathbf{A}_1\mathbf{T}\right)^k \right) \\
&= \mathbf{T}^T e^{-\mathbf{A}_3} e^{-\mathbf{A}_2}\mathbf{T} e^{-\mathbf{T}^T\mathbf{A}_1\mathbf{T}}.
\end{aligned} \tag{53}
$$

In the previous derivation, we used the relationship

$$
\mathbf{A}^k\mathbf{T} = \mathbf{T}\left(\mathbf{T}^T\mathbf{A}\mathbf{T}\right)^k, \tag{54}
$$

which can be obtained by doing the following iterative algebraic substitutions, $\mathbf{A}^{k-1}\mathbf{A}\mathbf{T} = \mathbf{A}^{k-1}\mathbf{T}\left(\mathbf{T}^T\mathbf{A}\mathbf{T}\right) = \mathbf{A}^{k-2}\mathbf{A}\mathbf{T}\left(\mathbf{T}^T\mathbf{A}\mathbf{T}\right) = \mathbf{A}^{k-2}\mathbf{T}\left(\mathbf{T}^T\mathbf{A}\mathbf{T}\right)^2 \ldots = \mathbf{A}^{k-p}\mathbf{T}\left(\mathbf{T}^T\mathbf{A}\mathbf{T}\right)^p$, until $p = k$ .

We can apply the same procedure to those matrix exponentials involving $\mathbf{A}_2$ and $\mathbf{A}_3$, respectively, and Eq. (53) becomes

$$
\begin{aligned}
\mathbf{T}^T e^{-\mathbf{A}_3} e^{-\mathbf{A}_2} e^{-\mathbf{A}_1}\mathbf{T} &= \mathbf{T}^T e^{-\mathbf{A}_3}\mathbf{T} e^{-\mathbf{T}^T\mathbf{A}_2\mathbf{T}} e^{-\mathbf{T}^T\mathbf{A}_1\mathbf{T}} \\
&= \mathbf{T}^T\mathbf{T} e^{-\mathbf{T}^T\mathbf{A}_3\mathbf{T}} e^{-\mathbf{T}^T\mathbf{A}_2\mathbf{T}} e^{-\mathbf{T}^T\mathbf{A}_1\mathbf{T}} \\
&= e^{-\mathbf{T}^T\mathbf{A}_3\mathbf{T}} e^{-\mathbf{T}^T\mathbf{A}_2\mathbf{T}} e^{-\mathbf{T}^T\mathbf{A}_1\mathbf{T}}.
\end{aligned} \tag{55}
$$

To obtain the expression in Eq. (28), we substituted matrices $\mathbf{A}_1 = \left(D\boldsymbol{\Lambda} + i\gamma G\mathbf{B}\right)\delta$, $\mathbf{A}_2 = D\left(\Delta - \delta\right)\boldsymbol{\Lambda}$, and $\mathbf{A}_3 = \left(D\boldsymbol{\Lambda} - i\gamma G\mathbf{B}\right)\delta$ in Eq. (55) with the matrices in the matrix exponential terms in Eq. (26).